\documentclass{ws-jai}
\usepackage[flushleft]{threeparttable}

\catchline{}{}{}{}{}

\begin{document}

\title{Pulsar Timing at the Deep Space Network}

\author{J. Kocz$^{1,2\dagger}$, W. Majid$^{2}$, L. White${^2}$, L. Snedeker${^2}$, M. Franco${^2}$}

\address{
$^1$ California Institute of Technology, 1200 E California Blvd, Pasadena, CA 91125 USA\\
$^2$ Jet Propulsion Laboratory, California Institute of Technology, Pasadena, CA 91109 USA
}

\maketitle
\corres{$^\dagger$Corresponding author}

\begin{history}
\received{(to be inserted by publisher)};
\revised{(to be inserted by publisher)};
\accepted{(to be inserted by publisher)};
\end{history}

\begin{abstract}
The 70-m DSN's Deep Space Station antenna 14 (DSS-14) at Goldstone has recently been outfitted with instrumentation to enable pulsar searching and timing operation. Systems capable of similar operations are undergoing installation at DSS-63, and are planned for DSS-43.  The Goldstone system is the first of these to become operational, with a 640~MHz bandwidth stretching from 1325-1965~MHz. Initial results from the pulsar timing pipeline show short-term residuals of $< 100$~ns for pulsar B1937+21. Commissioning observations at DSS-14 to obtain a baseline set of TOA measurements on several millisecond pulsars are currently underway.
\end{abstract}


\section{Introduction}

\begin{figure}
\begin{center}
\includegraphics[width=6cm]{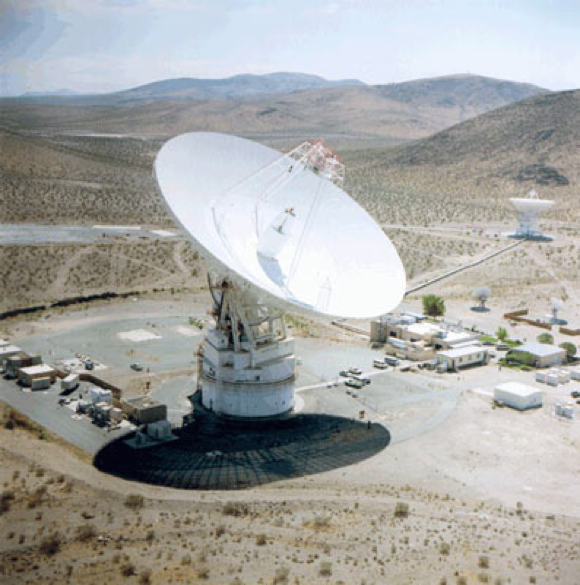}
\caption{DSS-14 at the Goldstone DSN site. \label{fig:dsn}}
\end{center}
\end{figure}

The Deep Space Network (DSN) is the spacecraft tracking and communication infrastructure for NASA's deep space missions.  It consists of three sites, approximately equally separated in terrestrial longitude, with multiple radio antennas at each site, which include a 70-m diameter antenna; Figure \ref{fig:dsn} shows one of these antennas.  We have recently installed and are currently commissioning a precision pulsar timing backend at DSS-14, the 70-m antenna at the Goldstone Deep Space Communications Complex (GDSCC).  The primary goal of this effort is to make use of DSN tracking schedule gaps to obtain precision time of arrival (TOA) measurements from an ensemble of millisecond pulsars (MSPs) for detection of gravitational waves.  Such pulsar timing array (PTA) measurements are sensitive to a range of gravitational wave (GW) frequencies around 1 nHz, determined by a combination of how frequently measurements are made and the span of the data sets \citep{hellings1983}.  Several PTAs are currently operational worldwide using some of the world's largest radio telescopes.  The International Pulsar Timing Array (IPTA) is a consortium of collaborations that includes NANOGrav \citep{jenet2009,mclaughlin2013}, the European Pulsar Timing Array (EPTA) \cite{kramer2013}, and the Parkes Pulsar Timing Array (PPTA) (Manchester, 2013; Manchester et al., 2013).  The development of this capability at the DSN gives the potential of high cadence TOA measurements in both the Northern and Southern hemispheres using theoretically identical receiving and processing chains. Once installed at all sites, testing will be undertaken to determine the extent to which having such similar systems can reduce the systematic errors that need to be determined and calibrated when using heterogenous systems. 

The DSN 70-m antennas are equipped with multiple receivers.  The primary receiver for pulsar timing measurements is the L-band system, spanning 640 MHz of bandwidth with a $T_{sys}$ of 25K.  While the current system at DSS-14 is capable of providing a single (left) circularly polarized channel, there is a funded effort at the DSN that is aiming to provide dual polarization (left and right circular) capability for both DSS-14 and DSS-43, the 70-m antenna at the Canberra Deep Space Communications Complex (CDSCC) in Australia. The DSS-14 antenna, as well as the other 70-m antennas in Canberra and Madrid, also have 2 and 8 GHz (dual polarization) receivers, with 90 and 500 MHz bandwidth respectively, which could be used to carry out pulsar timing measurements at higher radio frequencies. The pulsar timing system described here is based on the underlying design of the CASPSR\footnote{https://astronomy.swin.edu.au/pulsar/?topic=caspsr}  system implemented at the Parkes telescope. 

\section{Dataflow}
\label{sec:ware}

\begin{figure}
\begin{center}
\includegraphics[width=12cm]{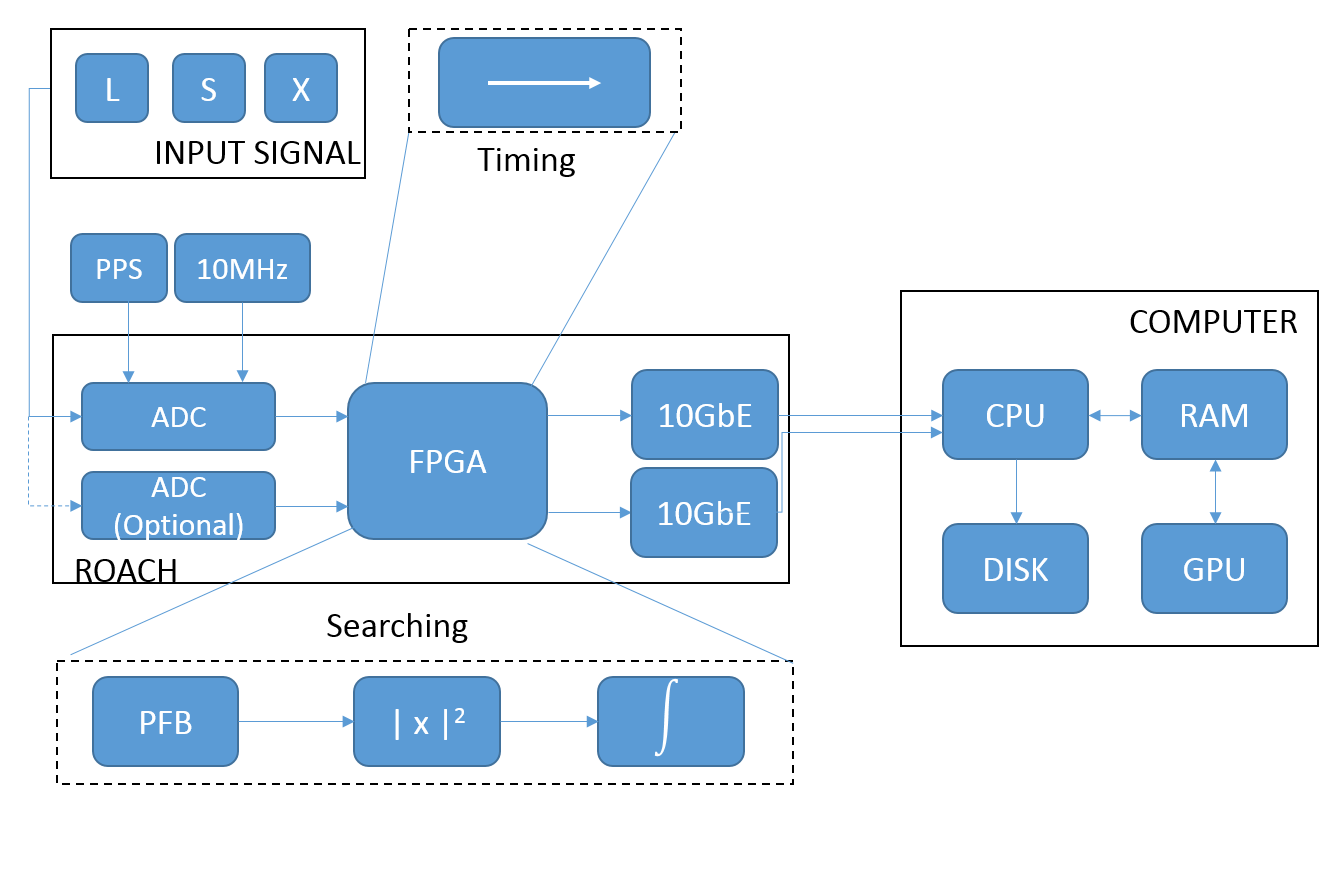}
\caption{The signal and dataflow path for the digital system. Data is received from either L, or S \& X band along with a 10MHz reference and PPS signal, and digitized with a 1280~MHz sampling rate, before being passed into an FPGA. Depending on the desired mode, the data is ``packetized'' for transmission over 10~GbE (timing) or passed through a polyphase filterbank, detected and accumulated, before being similarly read out over 10~GbE. The 10~GbE data is captured and placed into RAM, and then either written directly to disk for later offline processing (searching) or passed to a GPU for coherent dedispersion (timing). \label{fig:data_pipeline}}
\end{center}
\end{figure}

The system can be configured to run in either a timing (coherent) or searching (incoherent mode). The overall data flow for both configurations is shown in figure \ref{fig:data_pipeline}. While the hardware for both timing and searching pipelines are essentially the same, the firmware and processing software for the two tasks are quite different.  The pulsar equipment at Goldstone is ROACH\footnote{https://casper.berkeley.edu/wiki/ROACH} based, but has compatible ROACH2 based counterparts. The system also originally consisted of dual GTX690 GPUs, however, after failure one was replaced with a newer GTX980 model.

\begin{figure}
\begin{center}
\includegraphics[width=6cm]{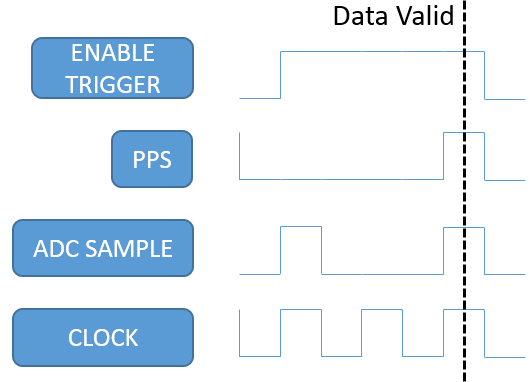}
\caption{When the observation is ready to start, the FPGA is told to ``arm'' and wait for the next PPS signal, which is locked to the observatory clock. The ADC data samples are considered valid from the ADC sample that aligns with the PPS.\label{fig:pps}}
\end{center}
\end{figure}

The signals are digitized using an 8-bit, dual 1.5~GSPS analog to digital converter (ADC) \footnote{see: KAT-ADC https://casper.berkeley.edu/wiki/KatADC}, sampling at 1280~MHz, which passes the data to a field programmable gate array (FPGA) for processing. In the processing of timing data, the FPGA plays only a small part. The commonly implemented CASPER methodology of a ``packetized'' system is used \citep{parsons2008}. The digitized inputs from the ADCs are read in $8\times$ parallel, and formed into 8192kB 10GbE UDP packets. The large packet size is chosen to minimize load on the CPU capture. The FPGA also accepts a pulse per second (PPS) synchronization pulse to initiate the data capture. The data capture is considered valid from the ADC sample corresponding to the PPS (figure \ref{fig:pps}). This ensures that the start time of the observation is accurate. 

In searching mode all of the initial required real time processing is performed on the FPGA, which acts as a standard digital spectrometer. After digitization, the signals are passed through a 2048-pt (1024 spectral channel) polyphase filterbank, detected and accumulated, before being transmitted via 10GbE to a computer host for storage. The accumulation time can be set by the user during runtime, and is limited only by the write speed of the disks on the storage computer. This data can then be processed offline by various pulsar software packages.

For timing, the high speed data capture and processing is performed by the use of \tt{PSRDADA} \rm{} \footnote{http://psrdada.sourceforge.net} and \tt{DSPSR}\rm{}\footnote{http://dspsr.sourceforge.net} \citep{vanstraten2011}. Data are captured on the 10GbE interface and stored in a ring buffer managed by \tt{PSRDADA} \rm{}. 
A well known and fundamental obstacle to pulsar timing is pulse dispersion. This occurs as lower frequencies of the broadband pulses from the pulsar are delayed more significantly by the interstellar medium than high frequency pulses. For processing reasons, pulsar timing and searching use different methods of compensating for this. For searching, the recorded filterbank data is dedispersed incoherently offline, where multiple dispersion measure trial values can be tested. For timing, coherent dedispersion \cite{hankins1975} as implemented on GPUs as part of \tt{DSPSR}\rm{}is used, with the processing spread across multiple GPUs.  In searching mode, the lower data rates are captured and written to disk by a simple socket interface implemented in C.  The run mode is selected and run primarily by script, with a simple GUI for initializing the observation which takes sources and integration times from a observation setup file.  The ADC data histogram can be read from the FPGA, and the gain adjusted over a 31.5~dB range as required. 

\section{First light}
\label{sec:first light}

\begin{figure}
\begin{center}
\includegraphics[width=9cm]{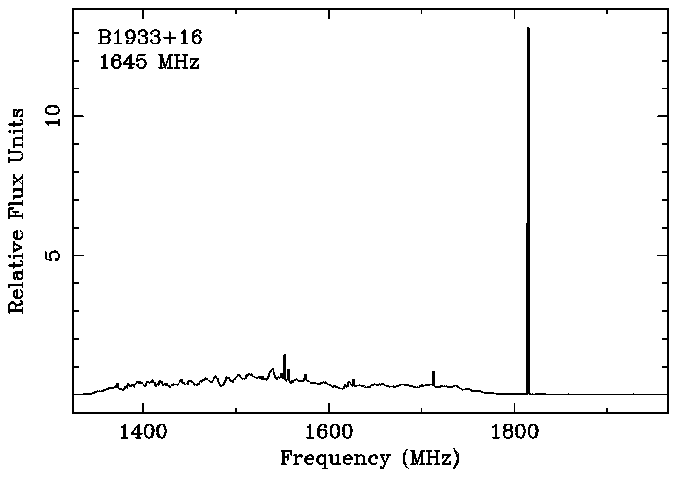}
\caption{An example spectrum, showing the strong RFI contamination in the bandpass, in particular the strong tone around 1815~MHz.  \label{fig:spectrum}}
\end{center}
\end{figure}

\begin{figure}
\begin{center}
\includegraphics[width=9cm]{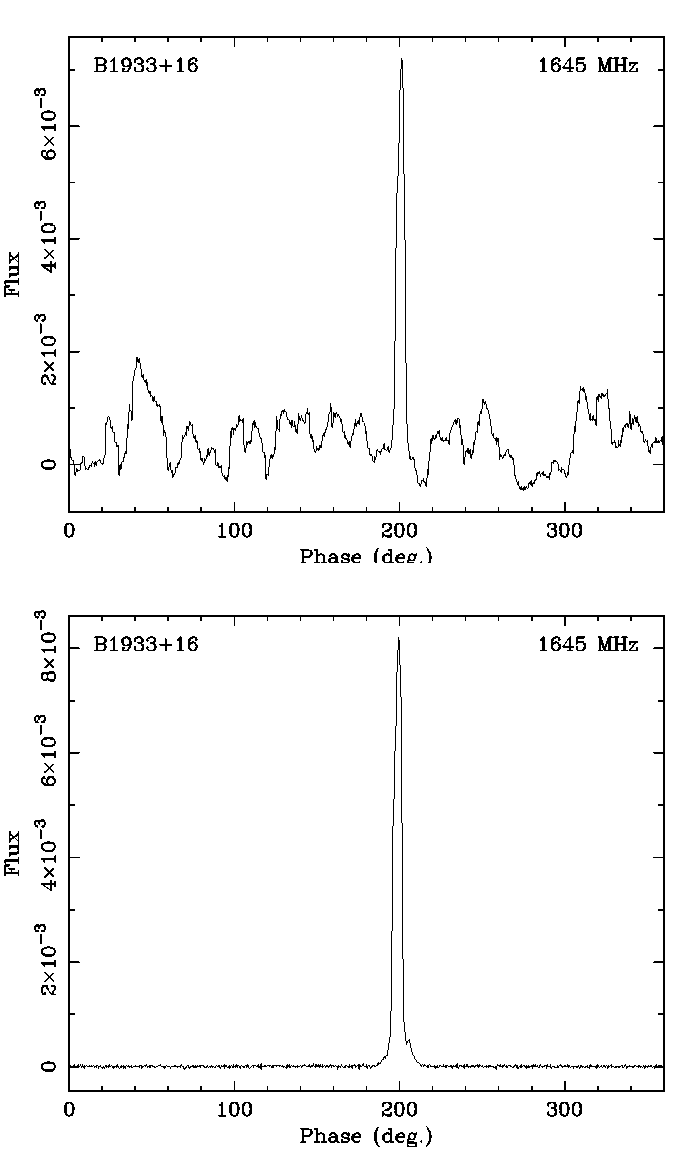}
\caption{Pulse profile of PSR B1933+16 before removal of RFI (top panel), and after (bottom). \label{fig:b1933_prof1}}
\end{center}
\end{figure}

\begin{figure}
\begin{center}
\includegraphics[width=12cm]{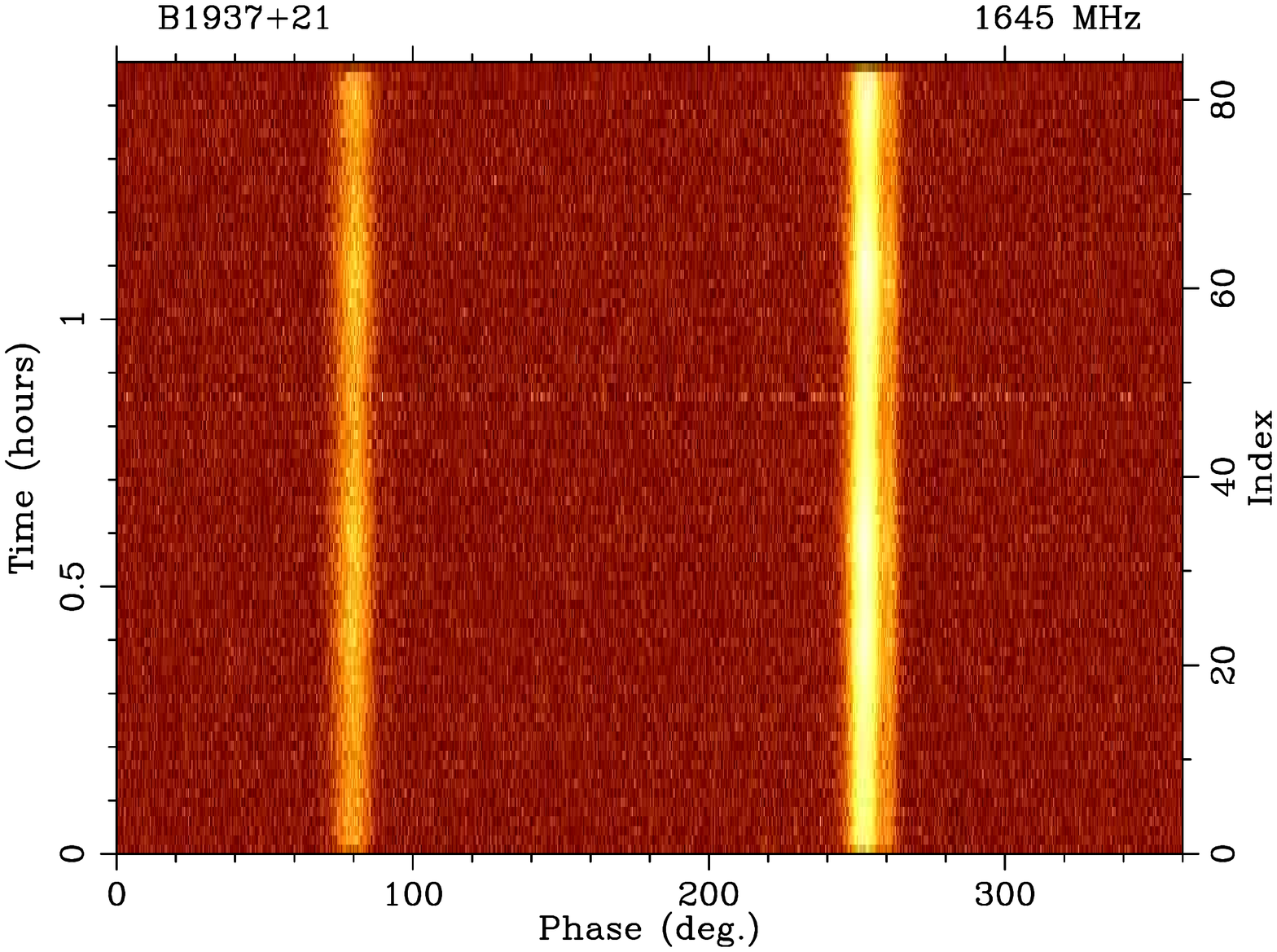}
\caption{An example time phase plot of the dedispersed profile for pulsar B1937+21, with a 1-min sub-integration time.\label{fig:pulsar}}
\end{center}
\end{figure}

Initial pulsar timing results have been obtained using DSS-14 at Goldstone. Upon analyzing the data however, one limitation discovered was the strong presence of radio frequency interference (RFI) in the L-band observations. The bandwidth at DSS-14 when using the L-band receiver is 640~MHz, stretching from 1325MHz - 1965MHz. As can be seen in Figure \ref{fig:spectrum}, some of the RFI at the higher frequencies was disruptive, dominating the signal power into the ADCs. A low pass filter was introduced at 1645~MHz in order to reduce the impact of this RFI contamination.  The two profiles in Figure \ref{fig:b1933_prof1} show a pulse profile of the bright pulsar PSR B1933+16 before and after removal of the RFI contaminated band, illustrating the vast improvement after the removal of the RFI contaminated band. The folded pulse profile and time vs frequency plots can also be monitored in quasi-realtime from the output data using the \tt{PSRCHIVE} \rm{}\footnote{http://psrchive.sourceforge.net} \cite{hotan2004} suite of tools (Figure \ref{fig:pulsar}). The results presented here are in arbitrary units. Flux calibration can be achieved using both a standard flux calibration source, and by injection of a noise diode signal.

\begin{figure}
\begin{center}
\includegraphics[width=10cm]{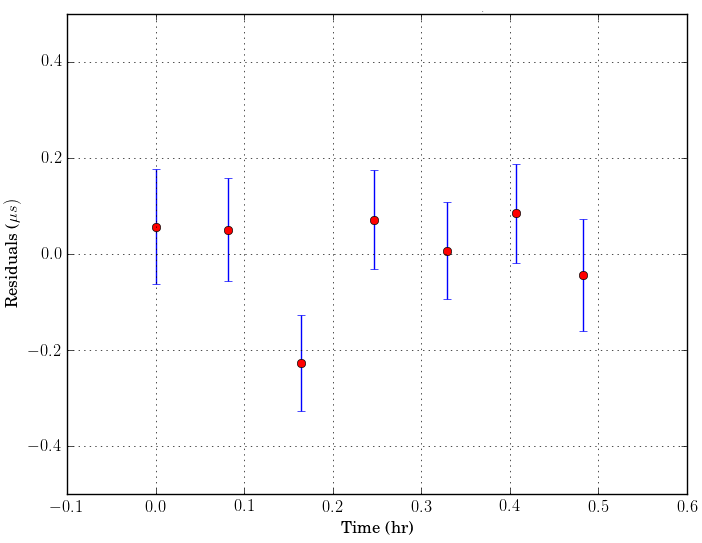}
\caption{TOA residuals from a single observation of PSR B1937+21.  \label{fig:toa1}}
\end{center}
\end{figure}

The first pulsar targeted testing of the timing system was B1927+21. A dedispersed pulse over a sample observation is shown in Figure \ref{fig:pulsar}, and initial time of arrival residuals in Figure \ref{fig:toa1}.  The TOAs were obtained by using the Nanograv template, and using the \tt{PAT} \rm{} tool in \tt{PSRCHIVE}\rm{} to cross correlate the measured profiles with the template. The Nanograv timing model for B1927+21 was then used to fit to the TOAs using the \tt{TEMPO2}\rm{} \cite{Hobbs2006} package. As only a small number of measurements have been taken so far, only the spin period and derivative were used as fitting parameters.  The results in figure \ref{fig:toa1} are based on a truncated passband after the installation of the low pass filter, resulting in 350~MHz of useful bandwidth.   The scatter in the residuals in this observations are below 100~ns, as expected given the detection sensitivity of the system.

\section{Conclusions and Future Directions}

We have described a pulsar timing system now installed at one of DSN's 70-m antennas in Goldstone, CA.  The system is capable of providing precision pulsar timing measurements at multiple frequency bands, though primarily used at L-band frequencies. 
Commissioning observations at DSS-14 to obtain a baseline set of TOA measurements on several millisecond pulsars are currently underway using gaps in the tracking schedule. The goal is to demonstrate the full capability of the system and carry out quantitative comparisons with TOAs obtained from other telescopes.  An upgrade to the DSS-14 system hardware in the form of increased GPU processing power is planned when the second polarization becomes available at the telescope. This will enable both polarizations to be processed. It is also planned to update the current RFI mitigating bandpass filter with notch filters as appropriate.  Future planned observing modes also include a real time single pulse search pipeline, operating both during dedicated time, and in a commensal mode during routine spacecraft communication.

\section*{Acknowledgments}
We thank Andrew Jameson and Willem van Straten for valuable discussion and debugging regarding adapting and using the PSRCHIVE, DSPSR and PSRDADA aspects of the system.
A portion of this research was performed at the Jet Propulsion Laboratory, California Institute of Technology, under a Research and Technology Development Grant and under a contract with the National Aeronautics and Space Administration.
Copyright 2016 California Institute of Technology. Government sponsorship acknowledged.

\clearpage

\end{document}